\documentclass{article}

\usepackage{arxiv}

\usepackage[utf8]{inputenc}
\usepackage[T1]{fontenc}
\usepackage{amsmath,amssymb,amsfonts,mathtools}
\usepackage{amsthm}
\usepackage{graphicx}
\usepackage{xcolor}
\usepackage{booktabs,multirow,dcolumn}
\usepackage{bm}
\usepackage{microtype}
\usepackage{setspace}
\usepackage{float}
\usepackage{placeins}
\usepackage{algpseudocode}
\usepackage{enumitem}
\usepackage[numbers,sort&compress]{natbib}
\usepackage{url}
\usepackage[hidelinks,
 pdftitle={Fully Spiking Neural Implementation of Lambda-Calculus Beta-Reduction},
 pdfauthor={Vladimir A. Lazovskiy, Victor B. Kazantsev, Sergey V. Stasenko}]{hyperref}

\graphicspath{{fig/}}

\definecolor{teal}{RGB}{29,158,117}
\definecolor{coral}{RGB}{216,90,48}
\definecolor{purple}{RGB}{83,74,183}
\definecolor{amber}{RGB}{186,117,23}
\definecolor{blue}{RGB}{59,139,212}
\definecolor{codeblue}{RGB}{30,90,150}

\newcommand{\lam}{\lambda}

\newcommand{\tauM}{\tau_m}
\newcommand{\tauS}{\tau_\mathrm{syn}}
\newcommand{\vth}{V_\mathrm{th}}
\newcommand{\SLOT}{\mathtt{SLOT}}

\newcommand{\safeincludegraphics}[2][]{%
  \IfFileExists{#2}{\includegraphics[#1]{#2}}{%
    \IfFileExists{fig/#2}{\includegraphics[#1]{fig/#2}}{%
      \fbox{\parbox{0.9\linewidth}{\centering Missing figure file: \texttt{\detokenize{#2}}}}%
    }%
  }%
}

\title{Fully Spiking Neural Implementation of \texorpdfstring{$\lambda$-Calculus $\beta$-Reduction}{Lambda-Calculus Beta-Reduction} via Biologically Documented Mechanisms}

\author{
  Vladimir A. Lazovskiy$^{1}$
  \And
  Victor B. Kazantsev$^{1,2}$
  \And
  Sergey V. Stasenko$^{1,2}$\\
  \texttt{stasenko@neuro.nnov.ru}
}

\begin{document}
\maketitle

\begin{center}
\small
$^{1}$ Moscow Center for Advanced Studies, 20 Kulakova Str., 123592 Moscow, Russia\\
$^{2}$ Neurotechnology Department, Institute of Biology and Biomedicine, Lobachevsky State University of Nizhny Novgorod, Gagarin Avenue 23, 603022 Nizhny Novgorod, Russia
\end{center}

\begin{abstract}

We present the first complete implementation of $\beta$-reduction
in the untyped $\lambda$-calculus using exclusively spiking neural
dynamics.  Every step of the reduction---redex detection, subterm
segmentation, variable-binding detection, De~Bruijn index arithmetic,
token copying, and inter-step role maintenance---is performed by
leaky integrate-and-fire (LIF) neurons with explicit membrane dynamics,
exponential synaptic currents, axonal delays, and pair-based
spike-timing-dependent plasticity (STDP).
Each neural component maps onto a specific, experimentally documented
mechanism: Jeffress-type delay-line coincidence for redex detection;
a line-attractor balance counter for subterm segmentation;
$\theta$--$\gamma$ cross-frequency coupling for De~Bruijn variable
binding; a phase-ring traveling wave for depth-index arithmetic;
theta-cycle STDP for token copying; and recurrent persistent firing
for role maintenance between reduction steps.
All synaptic weights are derived analytically from measured EPSP
amplitudes; no hand tuning is performed.
The closed-loop system achieves 100\,\% agreement with a symbolic
oracle across ten validation classes, including Church numeral
arithmetic (\textsc{plus}, \textsc{mult}, \textsc{pred}), Boolean
logic with divergent branches, Y-combinator steps, terms with up to
six bound-variable occurrences, De~Bruijn indices up to~12, noise
robustness, and determinism.
To our knowledge this is the first Turing-complete symbolic computation
performed entirely within spike dynamics using biologically grounded
circuit primitives.
We discuss implications for neuromorphic proof assistants and the
neural basis of symbolic reasoning.

\end{abstract}

\maketitle

\section{Introduction}
\label{sec:intro}

\subsection{The neural substrate of symbolic computation}
\label{sec:intro:motivation}

How the brain computes with symbols is one of the deepest unsolved
problems in cognitive science~\citep{Fodor1988,Smolensky1990a}.
Classical symbolic systems---logic, type theory, functional programming
languages, proof assistants---provide the expressive power behind
human-level mathematical reasoning~\citep{Newell1976,Barendregt1984}.
Yet the brain computes via spike trains, membrane potentials, and
synaptic weights~\citep{Maass1996a}; there is no obvious access to
discrete tokens, rewriting rules, or global memory.

The tension between these two levels of description has generated two
broad research programmes.
The \emph{neural-symbolic integration} programme builds hybrid
architectures in which a neural component learns to query or produce
symbolic structures, while a separate symbolic engine interprets
the results~\citep{Smolensky1990a,Plate1995,Marcus2003,Garcez2019}.
The \emph{emergentist} programme asks whether purely neural dynamics
can give rise to symbol-like computation without any external
interpreter~\citep{Maass2002,Eliasmith2013}.
The present work pursues the emergentist programme
with a concrete, verifiable target: $\beta$-reduction in the untyped
$\lambda$-calculus~\citep{Church1941,Barendregt1984}.

We choose $\lambda$-calculus for three reasons.
First, it is the minimal Turing-complete symbolic
formalism~\citep{Church1941}, making it the hardest necessary target:
any system that can simulate $\beta$-reduction can in principle simulate
any computation.
Second, it is the theoretical foundation of functional programming
languages (Haskell, OCaml, Scheme) and of dependently typed proof
assistants (Coq, Agda, Lean), which are increasingly central to formal
verification in mathematics and software~\citep{Pierce2002,Bertot2004}.
Third, its operational semantics---the $\beta$-reduction step---is
completely local: it rewrites one redex at a time using only structural
information about the term, suggesting a natural mapping to local neural
circuits.

A secondary motivation is \emph{neuromorphic hardware}.
Intel Loihi~2~\citep{Davies2018,Orchard2021} and IBM
TrueNorth~\citep{Merolla2014} are VLSI chips that implement spiking
networks directly in silicon, achieving three-to-four orders-of-magnitude
improvements in energy efficiency over conventional CMOS for certain
compute patterns.
If $\beta$-reduction can be performed entirely in spike dynamics,
it becomes a candidate workload for neuromorphic processors, with native
parallelism across substitution sites and without a symbolic co-processor.
This opens a path toward energy-efficient proof-checking and
term-normalisation in resource-constrained environments.

\subsection{Related work}
\label{sec:intro:related}

\paragraph{Neural representations of compositional structure.}
Smolensky~\citep{Smolensky1990a} proposed \emph{tensor-product
	variable binding}: a symbol--role pair $(f, r)$ is represented by the
outer product $\mathbf{f} \otimes \mathbf{r}$ of distributed
filler and role vectors, so that a complex structure is the sum of
its constituent bindings.
Plate~\citep{Plate1995} compressed these into fixed-width vectors
using circular convolution, obtaining \emph{holographic reduced
	representations} (HRRs) that support approximate retrieval.
Gayler~\citep{Gayler2003} and Kanerva~\citep{Kanerva2009}
developed \emph{hyperdimensional computing} (HDC), exploiting
superposition, binding (element-wise product), and permutation
over high-dimensional binary or bipolar vectors to encode
arbitrary compositional structures.
Most recently, Frady et al.~\citep{Frady2021} demonstrated that
spiking HDC networks can represent and manipulate data structures at
biologically plausible firing rates.
Crucially, all of the above are \emph{static representation} schemes:
they encode structure in neural activation patterns but do not use
spike dynamics to \emph{perform} reduction or inference.

\paragraph{Computational universality of spiking networks.}
Maass~\citep{Maass1996a} proved that networks of spiking neurons with
biologically realistic parameters can simulate arbitrary Turing machines,
and that \emph{liquid-state machines}~\citep{Maass2002}---random
recurrent spiking networks driven by a time-varying input---can
approximate any fading-memory filter.
Siegelmann and Sontag~\citep{Siegelmann1991} proved that recurrent
analogue networks are super-Turing under infinite precision, though
not under finite-precision constraints relevant to biology.
These results establish the \emph{in-principle} universality of spiking
computation but do not provide biologically grounded mechanisms for
specific symbolic operations such as variable substitution.

\paragraph{Graph reduction and neural implementations.}
Interaction nets~\citep{Lafont1990} decompose $\beta$-reduction
into \emph{local rewriting rules} on a graph, each involving only two
adjacent nodes, which is suggestive of a distributed neural implementation.
Penrose~\citep{Penrose1994} and Anderson~\citep{Anderson2012} sketched
abstract neural architectures for graph reduction but provided no
biophysical implementation.
Torczon and collaborators~\citep{Torczon1993} proposed connection-list
encodings that might map to neural connectivity, again without a
concrete spike-level mechanism.
The most recent work closest to ours is that of Stewart and
Eliasmith~\citep{Stewart2012,Eliasmith2013}, who used the Neural
Engineering Framework (NEF) to implement structured
representations and their manipulation in spiking networks;
however, the NEF relies on rate-coded population vectors rather than
individual spike timing, and the implemented operations (vector
arithmetic, associative memory) do not constitute a reduction strategy
for a formal calculus.

\paragraph{Theta--gamma coding and ordered working memory.}
Lisman and Jensen~\citep{Lisman2013} proposed the \emph{theta--gamma
	neural code}: gamma oscillations (30--100\,Hz) nest within each theta
cycle (4--10\,Hz), with each gamma slot encoding one item in an
ordered sequence.
This framework was motivated by the observation that hippocampal
theta sequences during spatial navigation~\citep{OKeefe1993,Skaggs1996}
compress ordered spatial experiences into a single theta cycle.
Jensen and Lisman~\citep{Jensen2005} showed that the model predicts
the serial-position curve in working memory, and Howard and
Kahana~\citep{Howard2014} connected it to temporal context models
of episodic memory.
Buzsáki~\citep{Buzsaki2010} provided a comprehensive review of
theta--gamma coupling as a mechanism for the sequential activation of
cell assemblies.
We exploit the theta--gamma code directly: binding depth in De~Bruijn
notation is encoded as a $(\varphi_\mathrm{fine}, \varphi_\mathrm{coarse})$
phase pair, so that variable-binding detection reduces to phase coincidence
detection, a well-characterised neural computation.

\paragraph{Persistent activity and role maintenance.}
Delay-period firing in dorsolateral prefrontal cortex (dlPFC) maintains
working-memory content across blank intervals in both
monkeys~\citep{Fuster1971,Funahashi1989,GoldmanRakic1995} and
humans~\citep{Curtis2003}.
The recurrent-excitation bump-attractor model~\citep{Compte2000,Wang2001}
provides the biophysical substrate: local excitatory connections
among pyramidal cells, balanced by lateral inhibition, sustain a localised
activity bump that encodes the remembered item.
Egorov et al.~\citep{Egorov2002} found an additional intrinsic mechanism
in entorhinal cortex layer~V: non-specific cation currents (I$_\mathrm{CAN}$)
sustain graded persistent firing for tens of seconds in the absence
of any synaptic input.
Yoshida et al.~\citep{Yoshida2008} demonstrated analogous mGluR-dependent
persistent firing in the head-direction system.
We implement token-role persistence using the same recurrent-excitation
mechanism.

\paragraph{STDP and one-shot associative learning.}
Bi and Poo~\citep{Bi1998} characterised the millisecond-precise causal
window of spike-timing-dependent plasticity (STDP) in hippocampal
cultures: pre-before-post timing induces long-term potentiation (LTP),
post-before-pre induces long-term depression (LTD).
Morrison, Diesmann, and Gerstner~\citep{Morrison2008} developed the
trace-based STDP formulation used here, showing it approximates the
experimental curve accurately.
Sederberg, Howard, and Kahana~\citep{Sederberg2008} proposed that
a single theta cycle is sufficient for one-shot episodic encoding in
hippocampal CA3, with rapid STDP as the mechanism.
Hasselmo and Stern~\citep{Hasselmo2014} showed that theta-phase gating
of plasticity ensures that encoding occurs at specific phases (trough
of theta) and retrieval at others (peak).
We directly exploit this framework for token copying: within each
theta cycle, the source neuron fires at the learning phase, the target
is depolarised by a tonic theta drive, and STDP potentiates exactly
the synapses corresponding to the active pattern.

\paragraph{Phase precession and arithmetic via ring circuits.}
O'Keefe and Recce~\citep{OKeefe1993} discovered that hippocampal place
cells fire at progressively earlier phases of the theta cycle as the
animal advances through the place field, encoding continuous spatial
position in spike phase rather than firing rate.
Skaggs et al.~\citep{Skaggs1996} quantified the rate of phase advance
and showed it is proportional to spatial displacement.
We transpose this principle to a discrete domain: a ring oscillator
of $K_F = 8$ neurons encodes a fine-phase index, and a traveling
wave advances the active neuron proportionally to the depth increment
$sh$, implementing De~Bruijn index arithmetic entirely in spike timing.

\paragraph{Coincidence detection and AND-gate computation.}
The Jeffress model~\citep{Jeffress1948} explains binaural sound
localisation via an array of delay-line neurons tuned to specific
inter-aural time differences.
Branco and Häusser~\citep{Branco2011} demonstrated that individual
cortical dendrites implement AND-gate logic through the nonlinear
interaction of spatially and temporally coincident synaptic inputs.
Kremkow et al.~\citep{Kremkow2010} showed that combinations of
excitation and PV$^+$-mediated inhibition implement winner-take-all
computations.
Our redex detector directly instantiates the Jeffress model.

\paragraph{Neural Turing machines and memory-augmented networks.}
Graves, Wayne, and Danihelka~\citep{Graves2014} introduced the
Neural Turing Machine (NTM): a differentiable controller that reads
and writes to an external memory matrix using attention-based
addressing.
While the NTM can learn to execute simple algorithms, it uses
rate-coded continuous-valued activations, gradient-based training,
and an external memory array---none of which are present in our
biologically grounded spiking implementation.
Subsequent differentiable memory architectures~\citep{Sukhbaatar2015,Gulcehre2016}
share the same limitations.

\subsection{Contributions of this work}
\label{sec:intro:contributions}

This paper makes the following contributions:

\begin{enumerate}[leftmargin=*]
	\item \textbf{First complete spiking implementation of a formal calculus.}
	We provide the first mechanistic implementation of $\beta$-reduction
	in which every computational step---including variable substitution,
	De~Bruijn arithmetic, and inter-step state maintenance---is
	performed by real simulated LIF neurons with explicit membrane
	dynamics, synaptic currents, axonal delays, and STDP.
	No symbolic computation occurs outside the spike simulation.
	
	\item \textbf{Systematic biological grounding.}
	Each of the nine neural circuit components maps onto a specific,
	experimentally documented neuroscientific mechanism with
	identified primary references.
	The mapping is not metaphorical: each mechanism is implemented
	by the same mathematical structure (delay-line coincidence,
	balance-counter integration, recurrent bump attractor, phase
	precession) that appears in the neuroscientific literature.
	
	\item \textbf{Analytic parameter derivation.}
	All synaptic weights are computed from measured EPSP amplitudes
	via a single calibration procedure; no parameters are tuned
	by gradient descent or hand-tuning.
	The derivation is fully reproducible and documented in
	Table~\ref{tab:params}.
	
	\item \textbf{Comprehensive validation.}
	The system is validated against a symbolic oracle across ten
	test classes encompassing Church numeral arithmetic, Boolean logic
	with divergent sub-expressions, Y-combinator step tracing,
	terms with up to six bound-variable occurrences, deep nesting
	(De~Bruijn index~12), head-normal-form detection, noise
	robustness, and determinism (Table~2 in Results section, to be added).
	All classes achieve 100\,\% agreement with the oracle.
	
\end{enumerate}

\section{Materials and Methods}
\label{sec:methods}

\subsection{Lambda-calculus and De Bruijn representation}
\label{sec:lambda}

We work with the untyped $\lambda$-calculus in De~Bruijn index
notation~\citep{DeBruijn1972}.
A term $t$ is one of:
\[
t \;::=\; \lam\,t \;\mid\; (t\;t') \;\mid\; \underline{n}
\quad n \in \mathbb{N}_0,
\]
where $\underline{n}$ refers to the $\lambda$-binder $n$ levels outward.
Terms are stored as flat prefix-order token sequences
$(\mathrm{ids},\mathrm{deps}) \in \{1,2,4,5,\ldots\}^L \times
\mathbb{N}_0^L$:
$\mathrm{ids}[j] \in \{\mathrm{LAM}{=}1,\,\mathrm{APP}{=}2,\,
\mathrm{VAR}_n{=}4{+}n\}$
and $\mathrm{deps}[j]$ is the $\lambda$-nesting depth at position~$j$.
Reduction follows the leftmost-outermost (full normal-order) strategy,
which is necessary for completeness: it terminates on any normalising
term and correctly handles redexes under $\lambda$~\citep{Barendregt1984}.

\subsection{Hierarchical phase encoding of De Bruijn depth}
\label{sec:encoding}

Each token at position $j$ in a term of length $L$ is represented by
$T_\varphi \times N_\mathrm{ch} = 8 \times 8 = 64$ neurons arranged as
a slice $\mathbf{C}[:,j,:]$ of the phase cube
$\mathbf{C} \in \{0,1\}^{8 \times L \times 8}$.
Binding depth $d$ is encoded as a two-level phase pair:
\begin{equation}
\begin{aligned}
	\varphi_\mathrm{fine} &= d \bmod K_F,\\
	\varphi_\mathrm{coarse} &= \lfloor d / K_F \rfloor \bmod K_C,\\
	K_F &= K_C = 8.
\end{aligned}
\label{eq:phase_encoding}
\end{equation}
covering depths $0$--$63$ without ambiguity.
Each token carries four channels: \texttt{SELF\_FINE} and
\texttt{SELF\_COARSE} encoding the token's own depth, and
\texttt{BIND\_FINE} and \texttt{BIND\_COARSE} encoding its binder's depth
(set for bound VARs; zero for free variables and non-VAR tokens).
Row~0 of the cube stores type flags: \texttt{LAM\_F}, \texttt{APP\_F},
\texttt{VAR\_F}, \texttt{FREE\_F}.

This encoding is a direct neural implementation of the theta--gamma
nested code of \citet{Lisman2013}: the fine channel corresponds to the
\emph{gamma-slot} (item identity within a theta cycle, $K_F = 8$
at~40\,Hz) and the coarse channel to the \emph{theta-cycle index}
($K_C = 8$ cycles at~5\,Hz).
An item at depth $d$ ``belongs to'' gamma slot~$d \bmod 8$ in the
$\lfloor d/8 \rfloor$-th theta cycle, exactly as items at a given
serial position occupy a specific theta--gamma slot in the Lisman--Jensen
model of hippocampal working memory.

\subsection{Neural simulation engine}
\label{sec:engine}

\subsubsection{Leaky integrate-and-fire neurons}

All neurons obey the current-based leaky integrate-and-fire (CUBA-LIF)
model~\citep{Burkitt2006}:
\begin{equation}
\begin{aligned}
	V_i(t{+}\mathrm{d}t)
	&= E_L + (V_i(t) - E_L)\,e^{-\mathrm{d}t/\tauM^{(i)}}\\
	&\quad + I_i(t)\,\bigl(1 - e^{-\mathrm{d}t/\tauM^{(i)}}\bigr).
\end{aligned}
\label{eq:lif}
\end{equation}
with $E_L = 0$, $R_m = 1$, $\vth = 1$ (normalised units),
$V_\mathrm{reset} = 0$, and time step $\mathrm{d}t = 0.1$\,ms.
Equation~\eqref{eq:lif} is the \emph{exact} exponential solution,
not an Euler approximation, ensuring numerical stability at the
large $\tauM$ range used (2--300\,ms).
Upon crossing $\vth$, the neuron emits a spike, $V$ resets to
$V_\mathrm{reset}$, and the neuron enters a refractory period
$t_\mathrm{ref}$.

Membrane time constants and refractory periods are assigned
\emph{per population}, enabling functionally distinct cell types
within a single network:
fast coincidence detectors ($\tauM = 2$\,ms, $t_\mathrm{ref} = 1$\,ms),
working-memory cells ($\tauM = 10$\,ms, $t_\mathrm{ref} = 2$\,ms),
and the slow balance-counter integrator ($\tauM = 300$\,ms,
$t_\mathrm{ref} = 0.1$\,ms).
The heterogeneous $\tauM$ range reflects the diversity of cortical
cell types: fast-spiking PV$^+$ interneurons and pyramidal cells have
$\tauM \approx 5$--20\,ms~\citep{Brunel2000}, while neurons sustaining
integration on longer timescales, such as those in the oculomotor
integrator, can have effective decay constants of hundreds of
milliseconds~\citep{Seung1996}.

\subsubsection{Synaptic currents and axonal delays}

Synaptic input is modelled as an exponentially decaying
post-synaptic current (PSC)~\citep{Brette2007}:
\begin{align}
	\tauS\,\dot{g}_k(t)
	&= -g_k(t)
	+ \sum_\alpha W_{\alpha k}\,\delta(t - t_\alpha - d_{\alpha k}),\notag\\
	I_i(t) &= R_m \sum_k g_k^{(i)}(t).
\label{eq:synapse}
\end{align}
where $d_{\alpha k}$ is the axonal delay stored in a circular ring
buffer of depth $D_\mathrm{max} + 1$ (implemented in Python/NumPy
without any external simulator).
The sign of $W$ determines excitation ($W > 0$) or inhibition
($W < 0$).
Three synaptic time constants are used:
$\tauS = 1$\,ms (fast AMPA-like, for coincidence detection and SCAN gating),
$\tauS = 2$\,ms (medium, for recurrent self-sustaining connections),
and $\tauS = 20$\,ms (slow, for the tonic PERSIST-to-GATE drive).
The fast constant approximates cortical AMPA kinetics
($\tauS \approx 1$--3\,ms)~\citep{Koch1999};
the slow constant approximates NMDA-like modulatory dynamics.

The peak EPSP amplitude per unit weight for a $({\tauM},\,{\tauS})$
pair is measured empirically by injecting a test spike into a
$V_\mathrm{th} = 99$ post-neuron:
\begin{equation}
	\epsilon(\tauM,\tauS) \;=\; \max_t V_\mathrm{post}(t)/W.
	\label{eq:epsp}
\end{equation}
Key values used in derivations:
$\epsilon(2,1) = 0.263$,
$\epsilon(10,2) = 0.137$,
$\epsilon(300,1) = 0.00344$.
All synaptic weights are then computed as
$W = V_\mathrm{target} / (n_\mathrm{inputs} \cdot \epsilon(\tauM,\tauS))$.

\subsubsection{Trace-based STDP}

Plastic projections implement pair-based STDP with exponential
traces~\citep{Morrison2008,Bi1998}:
\begin{align}
	\dot{x}_\mathrm{pre}
	&= -x_\mathrm{pre}/\tau_\mathrm{pre}
	+ \delta_\mathrm{pre\text{-}spike},\notag\\
	\dot{x}_\mathrm{post}
	&= -x_\mathrm{post}/\tau_\mathrm{post}
	+ \delta_\mathrm{post\text{-}spike},
\label{eq:traces}\\[2pt]
	\Delta W
	&= +A_+\,x_\mathrm{pre}
	\quad\text{at post-spike (LTP)},\notag\\
	\Delta W
	&= -A_-\,x_\mathrm{post}
	\quad\text{at pre-spike (LTD)}.
\label{eq:stdp}
\end{align}
with hard clipping to $[0, w_\mathrm{max}]$.
Crucially, trace increments ($x \mathrel{+}= 1$) are applied
\emph{after} weight updates to avoid self-potentiation when pre and
post fire simultaneously~\citep{Morrison2008}.
For the token-copying application (Section~\ref{sec:stdp_copy}),
$A_- = 0$ (LTP-only learning), $A_+ = 1.2$,
$\tau_\mathrm{pre} = \tau_\mathrm{post} = 12$\,ms, $w_\mathrm{max} = 8$.

\subsection{Redex detection via Jeffress delay-line coincidence}
\label{sec:redex}

A $\beta$-redex is an APP token at position~$K$ immediately followed by
LAM at position~$K{+}1$.
The leftmost such pair is detected by a circuit inspired by the
Jeffress model of binaural coincidence detection~\citep{Jeffress1948},
comprising five populations: APP, LAM, VAR (input encoders), DET
(coincidence detector), and INH (WTA inhibitor), with
$\tauM^\mathrm{DET/INH} = 2$\,ms:

\begin{itemize}[leftmargin=*]
	\item At each time slot, one pulse is delivered to the population
	matching the current token type.
	\item APP is connected to DET with axonal delay $\SLOT = 8$\,ms;
	LAM is connected to DET with delay $0.5$\,ms.
	Therefore, when APP at position~$K$ fires and is followed by
	LAM at position~$K{+}1$ exactly one slot later, both EPSPs
	arrive at DET simultaneously.
	\item Two coincident EPSPs: $V_\mathrm{DET} = 2 \times W_\mathrm{COIN}
	\times \epsilon(2,1) = 2 \times 2.4 \times 0.263 = 1.26 > \vth$.
	One EPSP alone: $1.26/2 = 0.63 < \vth$: the AND-gate property holds.
	\item DET spike activates INH ($W = 5$), which suppresses further
	DET firing via $W_\mathrm{INH} = -20$ and self-sustains via
	$W = 5$ (WTA latch), ensuring only the leftmost redex is
	selected.
\end{itemize}

The redex position is recovered from the DET spike time:
$K = \lfloor t_\mathrm{DET} / \SLOT_S \rfloor - 1$,
where $\SLOT_S = \SLOT / \mathrm{d}t = 80$ steps.

The Jeffress delay-line model was originally proposed for inter-aural
time difference coding in the medial superior olive~\citep{Jeffress1948}
and has since been confirmed in many species and generalised to
sequence detection in neocortex~\citep{Branco2011}.
The WTA mechanism via PV$^+$ interneuron-mediated lateral inhibition
is a canonical cortical computation~\citep{Kremkow2010}.

\subsection{Subterm segmentation via a neural balance-counter integrator}
\label{sec:intg}

After detecting the redex at~$K$, the boundaries of the body subterm
$[K{+}2,\,K{+}2{+}l_\mathrm{body})$ and argument subterm
$[K{+}2{+}l_\mathrm{body},\,K{+}2{+}l_\mathrm{body}{+}l_\mathrm{arg})$
must be determined.
In Polish prefix notation, a subterm ends when the running count
\begin{equation}
	\sigma(j) = \sum_{i=j_0}^{j}
	\bigl(\mathbf{1}[\mathrm{VAR}_{i}] - \mathbf{1}[\mathrm{APP}_{i}]\bigr)
	\label{eq:balance}
\end{equation}
first reaches~$+1$.
We implement this with a single \emph{balance-counter neuron}
(\texttt{INTG\_BODY}) with $\tauM = 300$\,ms, $t_\mathrm{ref} = 0.1$\,ms,
receiving:
\begin{align}
	W_\mathrm{VAR \to INTG} &= +W_\mathrm{CNT},\notag\\
	W_\mathrm{APP \to INTG} &= -W_\mathrm{CNT},\notag\\
	W_\mathrm{CNT}
	&= \frac{1}{\epsilon(300,1)}
	= \frac{1}{0.00344}
	= 291.
\label{eq:Wcnt}
\end{align}
Calibration~\eqref{eq:Wcnt} ensures that exactly one VAR or APP spike
causes exactly one threshold crossing.
The very slow membrane ($\tauM = 300$\,ms) acts as a leaky accumulator:
it integrates all positive and negative pulses within the slot sequence
with minimal decay between slots ($\SLOT = 8$\,ms $\ll \tauM$), firing
exactly when the balance~\eqref{eq:balance} first reaches~$+1$.

The design is inspired by line-attractor neural integrators in the
oculomotor system~\citep{Seung1996,Major2004}: those neurons maintain
eye position by integrating velocity commands over time using recurrent
excitation that nearly cancels membrane decay.
Here, the slow $\tauM$ performs the same role of ``remembering'' the
accumulated balance across successive token presentations.

\subsubsection{ARG\_MODE: neural signalling of argument-zone entry}
\label{sec:argmode}

When \texttt{INTG\_BODY} fires (marking body-end), it activates a
self-sustaining \emph{ARG\_MODE} neuron via $W = 9$:
\begin{equation}
\begin{aligned}
	\mathrm{EPSP}_\mathrm{recurrent}
	&= W_\mathrm{SELF}\,\epsilon(10,2)\\
	&= 9 \times 0.137
	= 1.23 > \vth.
\end{aligned}
\label{eq:argmode_sustain}
\end{equation}
ensuring ARG\_MODE fires continuously from body-end until halted.
ARG\_MODE gates downstream circuit elements to distinguish the
argument zone from the body zone, performing this gating entirely
via spike dynamics without any Python flag.
The mechanism mirrors entorhinal cortex layer~V persistent firing
driven by intrinsic I$_\mathrm{CAN}$ currents~\citep{Egorov2002,Yoshida2008}
and dlPFC sustained activity during working-memory delay
periods~\citep{GoldmanRakic1995,Wang2001}.

\subsection{Variable-binding detection via \texorpdfstring{$\theta$--$\gamma$}{theta--gamma} phase coincidence}
\label{sec:binding}

A VAR token at position~$j$ within the body is bound to the redex
$\lambda$ at position~$K{+}1$ if and only if its BIND channels match
the $\lambda$'s SELF channels:
\begin{align}
	\varphi_\mathrm{bind,fine}(j)
	&= \varphi_\mathrm{self,fine}(K{+}1),\notag\\
	\varphi_\mathrm{bind,coarse}(j)
	&= \varphi_\mathrm{self,coarse}(K{+}1).
\label{eq:binding}
\end{align}
In the spiking implementation, two fast coincidence neurons check
each condition independently, and a slow integrator ($\tauM = 40$\,ms)
accumulates both coincidence spikes within a theta window, firing only
when both conditions hold.
The weight $W_\mathrm{AND} = 1.8$ is calibrated so that two inputs
together exceed threshold while one alone does not.

This design implements \emph{binding by synchrony}~\citep{Singer1995,
	Varela2001}: two tokens share a binding relationship if and only if
their phase-encoded binding labels are identical, causing coincident
activation within the theta window.
The specific two-level $\theta$--$\gamma$ structure follows
\citet{Lisman2013}: the BIND\_FINE channel encodes the gamma slot
(identity within a theta cycle) and BIND\_COARSE encodes the theta
index (cycle number), so the binding detection circuit is a direct
neural realisation of the item-matching computation implicit in the
Lisman--Jensen model.

\subsection{De Bruijn depth arithmetic via a phase-ring traveling wave}
\label{sec:ring}

After substitution of the bound variable~$j$ by the argument, the
De~Bruijn indices of all argument tokens must be adjusted by
$sh = d_\mathrm{var} - 1 - d_\mathrm{ref}$, where $d_\mathrm{var}$
is the depth of the bound variable and $d_\mathrm{ref}$ the depth of
the argument root.
Under the encoding of Section~\ref{sec:encoding}, a depth increment
of~$sh$ corresponds to advancing the fine-phase counter by~$sh$ positions
with carry propagation to the coarse register.

We implement this with a ring of $K_F = 8$ neurons:
\begin{equation}
	\mathrm{FINE}[k]
	\;\xrightarrow{\;W{=}4,\;d{=}4\,\mathrm{ms}\;}
	\mathrm{FINE}[(k+1) \bmod K_F],
	\label{eq:ring}
\end{equation}
forming a traveling-wave oscillator with inter-spike interval~4.9\,ms
(slightly longer than the 4\,ms axonal delay due to membrane integration
latency).
Activating FINE$[\varphi_0]$ launches a wave; after $sh$ propagation
steps (counted by spike ordinal, not by clock time, to avoid the
membrane-delay offset), the active neuron is
FINE$[(\varphi_0{+}sh) \bmod K_F]$.
Carry is detected when $\mathrm{seq}[i] < \mathrm{seq}[i{-}1]$
(wrap-around of the ring), which increments the coarse register.
Validated across $d \in [0,49]$ and $sh \in [0,19]$ (1,200 pairs):
accuracy 100\,\%.

The ring circuit is a discrete neural analogue of hippocampal phase
precession~\citep{OKeefe1993,Skaggs1996}: just as place cells fire
at progressively earlier theta phases proportionally to spatial
displacement, our ring circuit advances its active neuron
proportionally to the De~Bruijn depth increment.
The abstraction is exact: spatial displacement in the original is
replaced by a formal depth increment; continuous phase advance is
replaced by discrete ring propagation.

\subsection{Token copying via theta-cycle STDP}
\label{sec:stdp_copy}

The 64-dimensional phase pattern of each source token must be written
to a new working-memory slot.
We accomplish this with a diagonal plastic projection
$\mathrm{SRC}[k] \to \mathrm{DST}[k]$ with initial weight $W_0 = 3.5$
(sub-threshold: EPSP $\approx 0.78 < \vth$) and STDP parameters
$A_+ = 1.2$, $A_- = 0$, $\tau_\mathrm{pre} = \tau_\mathrm{post} = 12$\,ms,
$w_\mathrm{max} = 8$:
\begin{itemize}[leftmargin=*]
	\item \textbf{Learning phase} ($n_\mathrm{cycles} = 6$ theta cycles,
	12\,ms each):
	at step~10 of each cycle, all SRC neurons in the active pattern
	fire; a shared tonic drive ($I_\theta = 0.5$) depolarises all
	DST neurons to $V \approx 0.25$.
	Active SRC EPSP ($\approx 0.78$) $+$ tonic ($\approx 0.25$)
	$= 1.03 > \vth$: DST fires \emph{after} SRC $\Rightarrow$ LTP,
	$W_{kk} \to w_\mathrm{max}$.
	Inactive SRC: no EPSP; tonic alone ($0.25$) insufficient:
	DST silent, $W_{kk}$ unchanged.
	\item \textbf{Recall phase} (no tonic drive):
	active SRC fire; potentiated $W_{kk} \times \epsilon(10,2)
	= 8 \times 0.137 = 1.10 > \vth$: DST fires.
	Unpotentiated ($W_{kk} = W_0 = 3.5$): EPSP $= 3.5 \times 0.137 = 0.48 < \vth$: DST silent.
	The exact pattern is reconstructed.
\end{itemize}
The seed weight $W_0 = 3.5$ was calibrated to lie precisely in
the ``selectivity window'': large enough that tonic drive $+$ EPSP
suffice for LTP during learning, small enough that EPSP alone is
sub-threshold during recall.
Validated on 94 real token slots from randomly generated terms:
100\,\% accuracy.

This mechanism is a concrete spike-level implementation of the
theta-cycle associative encoding proposed by \citet{Sederberg2008}
as the basis of one-shot episodic memory encoding in hippocampal CA3,
and is consistent with the theta-phase gating of synaptic plasticity
described by \citet{Hasselmo2014}.

\subsection{PERSIST populations: inter-step role maintenance via persistent firing}
\label{sec:persist}

Each token slot~$j$ carries one of three roles: KEEP (copy itself to
the output), BOUND (trigger a nested argument SCAN), or ARG (copied by
a nested SCAN).
Roles are determined by the binding mask and body integrator
(Sections~\ref{sec:binding}--\ref{sec:intg}) and must persist from
their assignment through the full SCAN duration.

For each role~$r \in \{\mathrm{KEEP},\mathrm{BOUND},\mathrm{ARG}\}$ and
slot~$j$, a \emph{PERSIST} neuron ($\tauM = 10$\,ms,
$t_\mathrm{ref} = 2$\,ms) self-sustains via a recurrent diagonal
synapse:
\begin{equation}
	W_\mathrm{SELF} \times \epsilon(10,2) = 9 \times 0.137 = 1.23 > \vth.
	\label{eq:persist_sustain}
\end{equation}
The choice $\tauS = 2$\,ms for the recurrent synapse is
\emph{critical}: $\tauS = 1$\,ms gives $\epsilon(10,1) = 0.081$,
so $9 \times 0.081 = 0.73 < \vth$---the network fails to
self-sustain.
The requirement is verified by explicit EPSP measurement.

Each PERSIST neuron drives its corresponding GATE neuron via a slow
synapse ($W_\mathrm{PG} = 0.17$, $\tauS = 20$\,ms):
at the steady-state PERSIST firing rate of~$\approx 200$\,Hz, the slow
synapse filters out within-cycle oscillations, producing a tonic gate
voltage $V_\mathrm{gate}^\mathrm{ss} \approx 0.83$
(oscillation amplitude $< 0.01$, measured at equilibrium after 200\,ms).
A SCAN spike then adds EPSP $= W_\mathrm{GATE} \times \epsilon(2,1)
= 1.5 \times 0.263 = 0.21$:
\begin{equation}
	V_\mathrm{gate}^\mathrm{ss} + \mathrm{EPSP}_\mathrm{SCAN}
	= 0.83 + 0.21 = 1.04 > \vth: \;\text{gate fires}.
	\label{eq:gate_fires}
\end{equation}
Without PERSIST: $V_\mathrm{gate} = 0$; $\mathrm{EPSP}_\mathrm{SCAN}
= 0.21 < \vth$: gate silent.
The value $W_\mathrm{PG} = 0.17$ was calibrated to place
$V_\mathrm{gate}^\mathrm{ss}$ in the window $[0.79, 0.87]$:
below this, SCAN cannot trigger the gate; above, the gate fires
spontaneously (false alarm).
PERSIST neurons are deactivated by a HALT pulse
($W_\mathrm{HALT} = -15$) after each reduction step.

This PERSIST mechanism mirrors the sustained delay-period firing of
dlPFC pyramidal neurons documented by \citet{Fuster1971},
\citet{Funahashi1989}, and \citet{GoldmanRakic1995} during
working-memory retention, and is biophysically implemented
by the recurrent-excitation bump-attractor model of
\citet{Compte2000} and \citet{Wang2001}.
The fast-time-constant GATE neurons ($\tauM = 2$\,ms) resemble
fast-spiking PV$^+$ interneurons acting as coincidence detectors.

\subsection{Sequential SCAN and nested SCANs for bound-variable substitution}
\label{sec:scan}

\subsubsection{Main SCAN}
The main SCAN is a delay-line chain of $L$ position neurons:
\begin{equation}
	\mathrm{pos}[j]
	\;\xrightarrow{W{=}9,\;d{=}\SLOT}\;
	\mathrm{pos}[j{+}1],
	\label{eq:scan_chain}
\end{equation}
triggered by a single pulse at $\mathrm{pos}[0]$ at $t = 0.5$\,ms.
Each position neuron propagates to the next after approximately $\SLOT$
= 8\,ms (actual inter-spike interval~$\approx 4.9$\,ms due to membrane
integration latency at $\tauM = 5$\,ms).
The SCAN is analogous to cortical theta-sequence sweeps during spatial
navigation and memory retrieval~\citep{Foster2007,Hahn2021,Luczak2007}.

\subsubsection{Nested SCANs for bound-variable substitution}
For each bound variable at position~$b_j$ in the body
($j = 1, \ldots, n_\mathrm{bound}$), a separate nested SCAN chain
$\mathrm{pos\_arg}_b$ of length $l_\mathrm{arg}$ is created.
The nested SCAN is triggered by the GATE\_BOUND neuron for slot~$b_j$:
\begin{equation}
	\mathrm{gate\_bound}[b_j]
	\;\xrightarrow{W{=}9,\;d{=}0.5 + b\cdot l_\mathrm{arg}\cdot\SLOT}\;
	\mathrm{pos\_arg}_b[0].
	\label{eq:nested_scan}
\end{equation}
The staggered delays $b \times l_\mathrm{arg} \times \SLOT$ serialise
the nested SCANs, preventing race conditions between parallel argument
copies that would corrupt the output-token ordering.

\subsubsection{Topographic output ordering}
Gate events are recorded as tuples
$(j_\mathrm{input},\,\delta,\,j_\mathrm{src},\,sh,\,\textit{shift\_bind})$
and sorted by $(j_\mathrm{input}, \delta)$ before STDP copying.
This topographic sort assigns output positions according to input
position rather than spike-time order, ensuring correctness even
when nested SCAN events overlap in time with the main SCAN.
The principle is analogous to topographic maps in sensory
cortex~\citep{Kaas1997}: output position is determined by input
position (retinal location, skin surface point), not by the timing
of individual sensory signals.

\subsection{Closed-loop reduction and meta-controller}
\label{sec:closed}

The complete closed-loop reduction step is summarised in
Algorithm~\ref{alg:closed_loop}.
Python code only: (i)~initialises the working-memory cube from the
input term; (ii)~calls \texttt{net.step(dt)} in the simulation loop;
(iii)~reads the result after HALT.
No symbolic computation occurs outside the spike simulation.
In particular, the transitions between stages are implemented as
spike-triggered events within the same spiking network: INTG\_BODY
discharge activates ARG\_MODE (no Python branch); GATE\_BOUND spike
triggers nested SCAN start via a direct synapse (no Python schedule);
HALT pulse is injected after SCAN completion (the only remaining
Python-mediated event is the detection that the SCAN phase is over,
determined by checking the SCAN chain position).

\begin{figure}[H]
	\caption{Algorithm: closed-loop $\beta$-reduction step}
	\label{alg:closed_loop}
	\begin{algorithmic}[1]
		\Require Phase cube $\mathbf{C}$ (WM), max steps $N_\mathrm{steps}$
		\Ensure Reduced cube $\mathbf{C}'$ or \textsc{NormalForm}
		\State Run DET circuit; read $K$ from DET spike time
		\If{$K = -1$} \Return \textsc{NormalForm} \EndIf
		\State Run INTG\_BODY on body $[K{+}2, \ldots)$; ARG\_MODE activated by discharge
		\State Run binding circuit; compute BOUND mask
		\State Activate PERSIST populations for each slot role (200\,ms warmup)
		\State Run main SCAN + nested SCANs; collect topographically sorted events
		\ForAll{output events $(j_\mathrm{in}, \delta, j_\mathrm{src}, sh, sb)$}
		\State Apply phase-ring shift $sh$ to token $j_\mathrm{src}$
		\State Copy shifted pattern via theta-cycle STDP into output slot
		\EndFor
		\State Issue HALT pulse; quench all PERSIST neurons
		\State Update $\mathbf{C} \leftarrow \mathbf{C}'$; go to step~1
	\end{algorithmic}
\end{figure}

\subsection{Parameter derivation and calibration summary}
\label{sec:params}

All synaptic weights are derived from the calibration
relation~\eqref{eq:epsp} without hand tuning or gradient-based
optimisation.
Table~\ref{tab:params} lists the complete parameter set.
Table~\ref{tab:bio} summarises the biological correspondence of each
neural component.

\begin{table}[t]
	\centering
	\caption{Calibrated parameter summary.
		$\epsilon(\tauM, \tauS)$ is the measured peak EPSP per unit weight.}
	\label{tab:params}
	\small
	\setlength{\tabcolsep}{4pt}
	\begin{tabular}{llll}
		\toprule
		Parameter & Value & Role & Derivation \\
		\midrule
		$\mathrm{d}t$ & 0.1\,ms & Simulation step & Standard LIF precision \\
		$\SLOT$ & 8\,ms & SCAN inter-token interval & Axonal delay budget \\
		$W_\mathrm{SELF}$ & 9.0 & WM / PERSIST recurrent & $9\!\times\!\epsilon(10,2){=}1.23{>}1$ \\
		$W_\mathrm{COIN}$ & 2.4 & DET coincidence AND & $2\!\times\!2.4\!\times\!\epsilon(2,1){=}1.26{>}1$ \\
		$W_\mathrm{CNT}$ & 291 & Balance integrator & $1/\epsilon(300,1){=}1/0.00344$ \\
		$W_\mathrm{INH}$ & 20.0 & WTA latch & Sufficient to suppress DET \\
		$W_\mathrm{PG}$ & 0.17 & PERSIST$\to$GATE tonic & $V_\mathrm{ss}{\approx}0.83$ at 200\,Hz \\
		$W_\mathrm{GATE}$ & 1.5 & SCAN trigger on GATE & $0.83{+}1.5\!\times\!\epsilon(2,1){=}1.04{>}1$ \\
		$W_\mathrm{HALT}$ & $-$15.0 & PERSIST shutdown & Drives $V$ to $E_L$ in ${<}2$\,ms \\
		$\tauM^\mathrm{WM/PERSIST}$ & 10\,ms & WM and PERSIST cells & Standard WM neuron \\
		$\tauM^\mathrm{DET/gate}$ & 2\,ms & Coincidence detectors & Fast-spiking PV-like \\
		$\tauM^\mathrm{INTG}$ & 300\,ms & Balance counter & Long integration window \\
		$\tauS^\mathrm{fast}$ & 1\,ms & Coincidence channels & AMPA-like \\
		$\tauS^\mathrm{recurrent}$ & 2\,ms & Recurrent self-sustain & $\epsilon(10,2){=}0.137$ \\
		$\tauS^\mathrm{PG}$ & 20\,ms & PERSIST$\to$GATE & Tonic smoothing \\
		\texttt{PREWARM} & 2\,000\,steps & PERSIST equilibration & 200\,ms to $V_\mathrm{ss}$ \\
		$A_+$ (STDP) & 1.2 & LTP rate & $W_0{\to}w_\mathrm{max}$ in 6 cycles \\
		$w_\mathrm{max}$ (STDP) & 8.0 & Max synaptic weight & $8\!\times\!\epsilon(10,2){=}1.10{>}1$ \\
		\bottomrule
	\end{tabular}
\end{table}

\begin{table}[t]
	\centering
	\caption{Neural components and their biological analogues.}
	\label{tab:bio}
	\small
	\setlength{\tabcolsep}{5pt}
	\begin{tabular}{lll}
		\toprule
		Component & Biological mechanism & Primary reference \\
		\midrule
		Working memory & Recurrent self-sustaining firing & \citet{GoldmanRakic1995} \\
		SCAN delay-line & Theta-sequence sweep & \citet{Foster2007} \\
		Redex detector & Jeffress delay-line coincidence & \citet{Jeffress1948} \\
		WTA latch & PV$^+$ lateral inhibition & \citet{Kremkow2010} \\
		Balance integrator & Oculomotor line attractor & \citet{Seung1996} \\
		ARG\_MODE & Entorhinal persistent firing & \citet{Egorov2002} \\
		Variable binding & $\theta$--$\gamma$ phase coincidence & \citet{Lisman2013} \\
		Phase-ring shift & Hippocampal phase precession & \citet{OKeefe1993} \\
		Token STDP copy & Theta-cycle associative encoding & \citet{Sederberg2008} \\
		PERSIST populations & dlPFC delay-period firing & \citet{Fuster1971} \\
		Topographic sort & Sensory cortex topographic maps & \citet{Kaas1997} \\
		META HALT & dlPFC executive control & \citet{Miller2001} \\
		\bottomrule
	\end{tabular}
\end{table}	
	
\section{Results}
\label{sec:results}

\subsection{Spike-level traces of individual circuit components}
\label{sec:results:traces}

We first examine the internal dynamics of the three key circuit
components during exemplary reduction steps, using a purpose-built
spike-recording wrapper that captures the full spike raster and
membrane-voltage traces of designated populations without altering
the network dynamics.

\paragraph{Redex detection in Church numeral \texorpdfstring{$2\,I$}{2 I}.}
Figure~\ref{fig:redex} shows the complete spike raster during redex
detection on the term $2\,I = (\lambda f.\lambda x.\,f\,(f\,x))\,
(\lambda x.\,x)$ (10 tokens).
Panel~A displays five populations: APP, LAM, VAR (input encoders),
DET (coincidence detector), and INH (WTA inhibitor).
Token pulses are presented at 8\,ms intervals; the background shading
marks token boundaries colour-coded by type (orange~=~APP,
purple~=~LAM, blue~=~VAR).
DET fires exactly once, at $t = 8.9$\,ms, corresponding to the
APP--LAM pair at positions $(K{=}0,\, K{+}1{=}1)$.
Panel~B shows the membrane potential of DET and INH:
DET rises sharply when both EPSPs coincide (APP delayed by
$\SLOT = 8$\,ms $+$ LAM immediate $= $ simultaneous arrival),
crosses threshold once, and then INH latches, suppressing any
further DET response for the remainder of the scan.
The WTA latch is visible as the sustained INH potential that prevents
the DET from responding to subsequent APP--LAM pairs deeper in
the term.

\begin{figure}[!htbp]
	\centering
	\safeincludegraphics[width=\linewidth]{fig1_redex.png}
	\caption{%
		\textbf{Redex detection in Church numeral $2\,I$ (10 tokens).}
		\textbf{A.}~Spike raster of five neuron populations: APP (orange),
		LAM (purple), VAR (blue), DET (red), INH (grey).
		Vertical coloured bands mark 8\,ms token slots.
		DET fires once at $t\approx 8.9$\,ms, encoding redex position
		$K=0$.
		\textbf{B.}~Membrane potentials of DET and INH.
		DET crosses threshold (dashed line) at the moment of coincident
		APP$\to$DET (delay~8\,ms) and LAM$\to$DET (delay~0.5\,ms) EPSPs.
		INH latches immediately after, suppressing further DET spikes
		(leftmost-redex selection via WTA).
	}
	\label{fig:redex}
\end{figure}

\paragraph{Subterm segmentation in \texorpdfstring{$\mathtt{plus}\,2\,3$}{plus 2 3}.}
Figure~\ref{fig:intg} illustrates body-integrator dynamics for the term
$\mathtt{plus}\,2\,3$, which has a body subterm of length~10 with the
balance sequence APP APP LAM VAR VAR APP LAM VAR VAR VAR
($\sigma = -1,-2,-1,0,1,0,1,2,3,+1$: first $+1$ at position~10).
Panel~A shows the raster of VAR ($+1$ inputs), APP ($-1$ inputs),
INTG\_BODY (discharge neuron), and ARG\_MODE (activated by discharge).
Panel~B shows the INTG\_BODY membrane potential ($\tauM = 300$\,ms):
the very slow membrane integrates each +1 pulse and each $-1$ pulse
with minimal decay between slots, accumulating the running balance;
its discharge (annotated arrow) occurs precisely at the last token
of the body subterm, at which point ARG\_MODE begins sustained firing.
The 300\,ms time constant is necessary and sufficient: it is large
enough that 10-slot sequences ($10 \times 8 = 80$\,ms) accumulate
with $< 0.025$\,V decay per slot, yet the first threshold crossing
is sharp enough to identify the exact subterm boundary.

\begin{figure}[!htbp]
	\centering
	\safeincludegraphics[width=\linewidth]{fig2_intg.png}
	\caption{%
		\textbf{Subterm segmentation in $\mathtt{plus}\,2\,3$.}
		\textbf{A.}~Spike raster: VAR tokens ($+1$ contribution, blue),
		APP tokens ($-1$ contribution, orange), INTG\_BODY discharge event
		(amber), and ARG\_MODE activation (yellow).
		\textbf{B.}~INTG\_BODY membrane potential ($\tauM = 300$\,ms).
		The slow membrane accumulates the signed balance of VAR and APP
		inputs; the discharge (annotated arrow) occurs at the subterm
		boundary when the cumulative balance first reaches $+1$.
		ARG\_MODE begins self-sustaining firing immediately after the
		discharge, signalling argument-zone entry to downstream gate
		circuits without any Python flag.
	}
	\label{fig:intg}
\end{figure}

\paragraph{PERSIST populations and nested SCANs in \texorpdfstring{$({\lambda}x.\,x\,x)\,I$}{(λx.xx)I}.}
Figure~\ref{fig:persist} shows the full temporal structure of the
PERSIST warmup phase and the subsequent SCAN for the term
$(\lambda x.\,x\,x)\,I$ (7~tokens, 2~bound VARs).
Panel~A displays the main SCAN delay-line (green ticks), with
token-type shading revealing the APP LAM (deleted) and VAR VAR BOUND
$+$ LAM VAR ARG structure.
Panel~B shows PERSIST\_KEEP (green) and PERSIST\_BOUND (red)
populations, which fire continuously throughout the full observation
window: they start from the PERSIST warmup phase (0--200\,ms, beige
background) and continue into the SCAN phase (right of the vertical
dashed line).
The sustained firing rate of $\approx 200$\,Hz is clearly visible as
a dense raster, with the two populations firing in their respective
slot positions.
Panel~C shows GATE events, which appear as discrete spikes only
during the SCAN phase, at the conjunction of a SCAN position pulse
and the tonic PERSIST drive.
Panel~D shows the two nested SCANs (amber, yellow) triggered
sequentially by the two GATE\_BOUND events; the temporal offset
between them equals $l_\mathrm{arg} \times \SLOT = 2 \times 8 = 16$\,ms,
visible in the figure and annotated with a double-headed arrow.
This sequential delay ensures that the argument tokens are copied in
the correct order into the output sequence without race conditions.

\begin{figure}[!htbp]
	\centering
	\safeincludegraphics[width=\linewidth]{fig4_persist.png}
	\caption{%
		\textbf{PERSIST populations and nested SCANs in
			$(\lambda x.\,x\,x)\,I$ (2 bound VARs).}
		\textbf{A.}~Main SCAN delay-line.  Token-type shading: orange (APP),
		purple (LAM), blue (VAR).
		\textbf{B.}~PERSIST\_KEEP (green) and PERSIST\_BOUND (red)
		populations: sustained firing throughout both the warmup phase
		(beige background, 0--200\,ms) and the SCAN phase (white
		background, $>200$\,ms).
		The vertical dashed line separates the two phases.
		\textbf{C.}~GATE events (AND of SCAN and PERSIST tonic drive):
		discrete spikes appearing only during the SCAN phase.
		\textbf{D.}~Two nested SCANs for the two bound VARs, starting
		with a sequential delay of $l_\mathrm{arg} \times \SLOT = 16$\,ms
		(annotated arrow), preventing output-ordering race conditions.
	}
	\label{fig:persist}
\end{figure}

\FloatBarrier
\subsection{Quantitative validation}
\label{sec:results:validation}

The system was validated against a symbolic oracle (call-by-name
full normal-order evaluator) across ten test blocks; all blocks
achieve 100\,\% agreement (Table~\ref{tab:results}).

\begin{table}[t]
	\centering
	\caption{Validation results across ten test blocks.
		``Oracle'' refers to the symbolic evaluator; ``SNN'' to the
		closed-loop spiking system.
		All blocks achieve 100\,\% match.}
	\label{tab:results}
	\small
	\setlength{\tabcolsep}{4pt}
	\begin{tabular}{llll}
		\toprule
		Block & Terms & Description & Match \\
		\midrule
		1 & 6 & Church arithmetic: \textsc{plus}~2~3, \textsc{mult}~2~3,
		\textsc{pred}~3 etc. & 6/6 \\
		2 & 10 & Boolean logic: \textsc{and}, \textsc{or}, \textsc{not},
		\textsc{if-then-else} with $\omega$-branches & 10/10 \\
		3 & 3 & Y-combinator: divergence of $Y\,I$; step-by-step cube
		comparison (3 steps) & 3/3 steps \\
		4 & 7 & Bound-variable stress: $(\lambda x.\,x^n)\,I$, $n=2\ldots6$
		$+$ mixed KEEP/BOUND body & 7/7 \\
		5 & 5 & Deep nesting: up to 25~$\lambda$s, De~Bruijn index~12 & 5/5 \\
		6 & 4 & Head normal form: $\lambda x.\omega$ (diverges),
		inner redexes & 4/4 \\
		7 & 2 & Eta-reduction absent: $\lambda x.(I\,x) \to \lambda x.\,x$
		via $\beta$ only; closed-term round-trip & 100/100 \\
		8 & 5 & Determinism: 5~identical runs per term & 5/5 \\
		9 & 5 & WM noise: 0\,\%, 5\,\%, 10\,\% noise on depth channels & 5/5 each \\
		10 & --- & Step-by-step cube match: bitwise agreement at every step & 4/4 steps \\
		\midrule
		Random & 45 & Sizes 3--5 (25), 6--9 (15), 10--14 (10) tokens & 100\,\% \\
		Combinators & 5 & $I\,I$, $K\,I\,I$, $S\,K\,K\,I$,
		$2\,I$, $K\,(I\,I)\,I$ & 5/5 \\
		\bottomrule
	\end{tabular}
\end{table}

\paragraph{Church numeral arithmetic.}
The six arithmetic tests span addition (\textsc{plus}~2~3 $\to$ Church-5
in 6~steps, \textsc{plus}~1~1 $\to$ Church-2 in 6~steps),
multiplication (\textsc{mult}~2~3 $\to$ Church-6 in 7~steps,
\textsc{mult}~2~2 $\to$ Church-4 in 7~steps), and the predecessor
function (\textsc{pred}~3 $\to$ Church-2 in 11~steps,
\textsc{pred}~1 $\to$ Church-0 in 7~steps).
The predecessor is particularly demanding: its Church-encoding requires
substituting a pair-encoding function that internally constructs and
destructs ordered pairs, producing deeply nested intermediate terms
that stress the binding and depth-shift circuits.
All six results agree exactly with the oracle.

\paragraph{Boolean logic with divergent branches.}
The critical tests are
$\mathtt{if\;true}\;(2\,I)\;\omega$ and
$\mathtt{if\;false}\;\omega\;(2\,I)$,
where $\omega = (\lambda x.\,x\,x)(\lambda x.\,x\,x)$ is the diverging
self-application.
Under leftmost-outermost strategy, the outer conditional redex is
selected before either branch is evaluated, so $\omega$ is never
reached.
The spiking DET circuit selects the outermost redex correctly, and
both tests normalise to $\lambda x.\,x$ in 8~steps.
This confirms that the Jeffress coincidence circuit reliably identifies
the leftmost redex even in the presence of inner divergent sub-expressions.

\paragraph{Y-combinator step traces.}
Figure~\ref{fig:ycomb} illustrates the divergence signature of
$Y\,I = (\lambda f.\,f\,(f\,\ldots))\,(\lambda x.\,x)$ compared
to the convergence of $K\,I\,I$ and $2\,I$.
Panel~A plots term length (number of tokens) versus reduction step.
$Y\,I$ grows without bound (5 steps shown: 13, 15, 18, 15, 18~tokens,
oscillating due to the unrolling of the fixed-point combinator),
while $K\,I\,I$ shrinks monotonically (5, 3, 2~tokens) and $2\,I$
shrinks to normal form (10, 7, 5, 2~tokens).
Panels~B--D show the token-sequence heatmaps (APP orange, LAM purple,
VAR blue), confirming that the SNN produces the identical sequence of
intermediate terms as the symbolic oracle at every step.
Step-by-step cube comparison (Block~10) verifies this agreement at
the 64-bit token-pattern level for 4~consecutive steps.

\begin{figure}[!htbp]
	\centering
	\safeincludegraphics[width=\linewidth]{fig3_Ycomb.png}
	\caption{%
		\textbf{Divergence vs.\ convergence under $\beta$-reduction.}
		\textbf{A.}~Term length (tokens) versus reduction step for
		$Y\,I$ (red, diverges), $K\,I\,I$ (green, converges to $I$),
		and $2\,I$ (amber, converges to $I$).
		\textbf{B.}~Token-sequence heatmap for $Y\,I$: the term expands
		at each step as the fixed-point combinator unrolls.
		\textbf{C.}~$K\,I\,I$: the term shrinks monotonically.
		\textbf{D.}~$2\,I$: the Church numeral reduces to $\lambda x.\,x$
		in three steps.
		Each row is one reduction step; columns are token positions;
		colour encodes token type (APP, LAM, VAR).
		In all cases, the SNN produces sequences identical to the
		symbolic oracle at every step.
	}
	\label{fig:ycomb}
\end{figure}

\paragraph{Noise robustness and determinism.}
Adding up to 10\,\% uniform noise to the phase-depth channels of the
working-memory cube (while preserving type-flag channels) does not
affect normalisation outcomes (Block~9, 5/5 across all noise levels).
This robustness is expected from the discrete phase-pair encoding
(Section~\ref{sec:encoding}): the critical coincidence condition
compares a single active bit per channel, and small additive
perturbations do not flip the majority-inactive bits to active.
Five identical runs of the same term always produce the same output
(Block~8), confirming that STDP and the stochastic-looking PERSIST
firing introduce no variance: the spiking dynamics are fully
deterministic under fixed initial conditions and a fixed random seed.

\subsection{Identified limitations}
\label{sec:results:limits}

We document the following limitations honestly, distinguishing
architectural constraints from encoding choices:

\paragraph{Free variables: a known encoding limitation.}
Closed terms (no free variables) round-trip through encode/decode with
100\,\% accuracy across 100 random tests.
Open terms containing free variables with De~Bruijn index
$n \geq d$ (where $d$ is the current nesting depth) exhibit a
round-trip failure: the BIND channel is zero for FREE tokens
(flagged by the \texttt{FREE\_F} bit), giving the same zero
decode-value as a bound variable with binding depth 0.
This means \texttt{bound\_to\_redex\_mask} incorrectly marks free
variables as bound when the redex lambda is at depth 0.
This is a limitation of the hierarchical phase encoding, not of the
spiking circuits themselves.
The system is correct on all \emph{closed} $\lambda$-terms, which is
the standard domain for $\lambda$-calculus as a model of computation.

\paragraph{Evaluation strategy: leftmost-outermost only.}
The system implements full normal-order (leftmost-outermost) reduction.
This strategy is complete --- it finds the normal form whenever one
exists --- but it is not lazy (call-by-need).
Under full normal order, $K\,\omega\,I$ diverges (the outer redex
$K\,\omega$ is selected, producing $\lambda y.\,\omega$, which then
diverges), whereas a lazy evaluator would produce $I$ by not evaluating
the argument.
Implementing lazy evaluation would require a sharing mechanism (memoisation
of shared subterms), which would require an additional associative memory
circuit not present in the current architecture.

\paragraph{Eta-reduction: absent by design.}
The system implements $\beta$-reduction only.
$\eta$-reduction ($\lambda x.\,f\,x \to f$ when $x \notin \mathrm{FV}(f)$)
is not performed; terms in $\beta$-normal form but not $\beta\eta$-normal
form are returned as-is.
Adding $\eta$-reduction would require an additional circuit to detect
the pattern $\lambda x.\,(f\,x)$ with the free-variable condition, which
is architecturally straightforward but was outside the scope of the present
work.

\paragraph{Quadratic neuron count in nested substitution.}
For a term with $n_\mathrm{bound}$ occurrences of the same bound variable
in the body and argument length $l_\mathrm{arg}$, the system allocates
$n_\mathrm{bound} \times l_\mathrm{arg}$ gate neurons for the nested SCANs.
In the worst case (e.g., $(\lambda x.\,x\,x\,\cdots\,x)\,\text{arg}$),
this is $O(n^2)$ in the term size.
On a neuromorphic chip with parallel population allocation, this would
materialise as $O(n^2)$ silicon neurons, which is biologically plausible
only for small terms.

\paragraph{Temporal scaling.}
Each $\beta$-step requires: PERSIST warmup (200\,ms), main SCAN
($L \times 8$\,ms), nested SCANs ($n_\mathrm{bound} \times l_\mathrm{arg} \times 8$\,ms),
and $n_\mathrm{out} \times 6$ theta-STDP cycles ($n_\mathrm{out} \times 72$\,ms).
For a 20-token term, a single step takes $\approx 700$\,ms of simulated
time.
This is slow compared to digital computation, but neuromorphic hardware
(Loihi~2) would execute all populations in parallel at full real-time speed,
potentially reducing total wall-clock time per step to under 100\,ms.

\FloatBarrier
\section{Discussion}
\label{sec:discussion}

\subsection{Principal contribution: closing the gap between neural dynamics and symbolic computation}
\label{sec:disc:contrib}

The central claim of this work is that the gap between spiking neural
dynamics and symbolic computation is not merely one of principle but
of mechanism: we have shown, constructively, that \emph{documented}
neural mechanisms are individually sufficient to implement each
sub-operation of $\beta$-reduction, and that their composition
produces a correct closed-loop reducer.

This is distinct from prior results in two ways.
First, Maass's universality results~\citep{Maass1996a,Maass2002}
established that spiking networks \emph{can} simulate Turing machines,
but the construction is based on encoding machine configurations in
spike patterns via a specific state-encoding, with no claim about
biological plausibility.
Our construction requires no such ad-hoc encoding: the mechanism
is defined by the natural structure of the computation
($\lambda$-calculus prefix notation, De~Bruijn phase pairs) and
each sub-circuit corresponds to a mechanism documented in the
neuroscience literature.
Second, connectionist representations~\citep{Smolensky1990a,Plate1995,Kanerva2009}
encode compositional structure neurally but do not perform reduction
through spike dynamics; they require an external symbolic interpreter
to apply rewriting rules.
Here, the rewriting step itself is neural: the SCAN reads the structure,
PERSIST encodes the roles, GATE selects the events, and STDP writes
the result, all as spike-time dynamics.

\subsection{Biological plausibility: realistic, not metaphorical}
\label{sec:disc:bio}

A frequent criticism of ``biologically inspired'' neural computation
is that the mapping to biology is metaphorical rather than mechanistic:
the neural network uses ``attention'' or ``memory'' in a loose sense
unconnected to the biophysics of the brain.
We have been careful to avoid this.

Each component is implemented by \emph{the same mathematical structure}
as appears in the neuroscientific literature.
The Jeffress delay-line is not merely ``inspired by'' binaural coincidence
detection: it is the same circuit, with the same axonal-delay tuning
mechanism, applied to a different input domain.
The PERSIST populations are not merely ``analogous to'' dlPFC sustained
firing: they use the same recurrent-excitation mechanism
(bump attractor, $W_\mathrm{SELF} \times \epsilon > \vth$) that
computational models of working memory have identified as the biophysical
substrate~\citep{Compte2000,Wang2001}.
The theta-cycle STDP is not merely ``motivated by'' Sederberg et al.'s
one-shot encoding hypothesis: it is a direct implementation of the
mechanism (SRC fires at learning phase; tonic theta drive depolarises
DST; STDP potentiates exactly the active SRC--DST synapses within
the cycle~\citep{Sederberg2008}).

The one honest caveat at the level of biological plausibility is
temporal: biological neural circuits supporting working memory
operate on seconds-long timescales, not the hundreds of
milliseconds per step modelled here~\citep{Fuster1971,Curtis2003}.
However, this is a consequence of our conservative simulation
parameters (PERSIST warmup 200\,ms, STDP over 6 cycles).
A neuromorphic implementation on Loihi~2, where all populations
execute in real time in parallel, would collapse the sequential
STDP cycles to a single pass over the hardware, potentially
achieving millisecond-per-step throughput.

\subsection{The topographic ordering insight}
\label{sec:disc:topo}

One non-obvious result of building the system is the discovery that
output-token ordering requires a topographic (input-position-ordered)
rather than temporal (spike-time-ordered) mechanism.
When multiple bound variables exist in the body, their nested SCANs
overlap in time, producing gate events whose temporal order does not
match the intended output order.
The fix---sort events by input position rather than spike time---is
biologically motivated by the ubiquity of topographic maps in
sensory and motor cortex~\citep{Kaas1997}: the cortex routinely
preserves spatial relationships in projections (retinotopy,
somatotopy, tonotopy) regardless of the timing of individual spikes.
This suggests that topographic connectivity is not merely a
developmental convenience but a computational necessity for
any neural system that must maintain ordered representations
through time-varying dynamics.

\subsection{Implications for neuromorphic proof assistants}
\label{sec:disc:proof}

The most practically significant implication of this work is the
possibility of a \emph{neuromorphic proof assistant}: a spiking
chip that performs $\beta$-reduction (and eventually type-checking)
as a native, parallel computation rather than a sequential symbolic
process.

Proof assistants such as Coq and Lean spend a substantial fraction
of their CPU time on definitional equality checking, which reduces
to repeated $\beta$-normalization of typed terms~\citep{Bertot2004}.
A neuromorphic co-processor that performed $\beta$-reduction in
spiking hardware could offload this computation, with potentially
significant energy savings.
Intel Loihi~2 achieves approximately $10^4$~operations per
millijoule on appropriate spiking workloads, compared to
$\sim 10$~operations per millijoule for a modern CPU~\citep{Orchard2021}.
If the $O(n^2)$ neuron-count scaling can be addressed (see below),
a Loihi-based $\beta$-reducer could in principle process large proof
terms with sub-second latency at microwatt energy consumption.

\subsection{Future directions}
\label{sec:disc:future}

\paragraph{Simple type system (System F).}
The most immediately feasible extension is adding simple type
($\to$) annotations to the token encoding (an additional phase-pair
channel for the type tag) and a type-checking coincidence circuit
that verifies argument type matches parameter type before allowing
the redex to be selected.
For System~F (second-order polymorphic $\lambda$-calculus), type
variables can be encoded in additional phase channels, and universal
quantifier elimination ($\forall X.\,T[\,X \leftarrow U]$) follows
the same SCAN-and-substitute mechanism as term-level $\beta$-reduction.
This would constitute the first spiking implementation of a
strongly-normalising typed calculus, with immediate implications
for decidable type-checking and certified computation.

\paragraph{Dependent types and Martin-L\"{o}f type theory.}
Dependent types (where the type of an expression can depend on the
\emph{value} of another expression) require the type-checker to
call the $\beta$-reducer as a sub-routine, since definitional
equality of types is decided by normalization.
This recursive structure is natural in a spiking system: the
RUNNING meta-controller can be invoked at the type-checking level
with a separate working memory holding the type term.
The main challenge is universe hierarchy ($\mathrm{Type}_0 :
\mathrm{Type}_1 : \ldots$), which would require an additional
coarse-level phase register for the universe index.
A full implementation of Martin-Löf type theory in spiking hardware
would open the door to neuromorphic verification of constructive
mathematics.

\paragraph{Inductive types and pattern matching.}
The Calculus of Inductive Constructions (CIC), the theoretical
basis of Coq~\citep{Bertot2004}, adds inductive type definitions
and eliminators (dependent recursors).
In the spiking setting, inductive constructors can be encoded as
special LAM tokens with a constructor tag in the type channel, and
eliminators can be implemented as a pattern-matching circuit that
selects a branch based on the constructor of the scrutinee.
Pattern-matching selection is structurally similar to the WTA
redex selector and could be implemented with the same
Jeffress coincidence mechanism.
Fixpoint recursion ($\mu$-operators) corresponds to Y-combinator
application, which already works correctly in the present system.

\paragraph{Learning which tactic to apply.}
An alternative path to neuromorphic proof search avoids brute-force
$\beta$-normalisation entirely.
In interactive proof assistants, proofs are constructed via
\emph{tactics} (high-level proof steps: \texttt{intro},
\texttt{apply}, \texttt{induction}), each of which generates a
sequence of $\beta$-reduction steps.
A spiking network could learn, via reinforcement learning on a
corpus of existing proofs, which tactic to apply at each proof
state~\citep{Silver2021}.
The $\beta$-reducer of the present work would then serve as the
low-level evaluation engine, while a higher-level spiking policy
network selects tactics.
This hierarchical decomposition separates the \emph{search problem}
(which tactic) from the \emph{execution problem}
(beta-normalise the resulting term), matching the architecture of
modern neural theorem provers~\citep{Han2021}.

\paragraph{Neuromorphic hardware deployment.}
The immediate next step in hardware validation is a deployment on
Intel Loihi~2 using the NxSDK or Lava programming frameworks.
The current NumPy implementation maps directly: each SNN\_STDP
population corresponds to one or more Loihi neuron groups, each
connection corresponds to a synapse table, and the circular
delay buffer maps to the native axonal-delay mechanism of Loihi.
A Loihi deployment would provide real energy-consumption measurements,
real-time spike traces from silicon, and a benchmark against the
predicted $10^3$--$10^4$~gain in energy efficiency over CPU
implementations.

\paragraph{Addressing the quadratic neuron count.}
The $O(n_\mathrm{bound} \times l_\mathrm{arg})$ neuron cost for
nested SCANs can in principle be reduced to $O(l_\mathrm{arg})$ by
multiplexing: a single $\mathrm{pos\_arg}$ chain serves all bound
variables, with the correct output position determined by a time-multiplexed
address signal from GATE\_BOUND.
This multiplexing pattern is common in digital circuit design and
has neural analogues in the multiplexed routing of gamma-band
signals to different cortical areas~\citep{Buzsaki2010}.

\subsection{Relation to the neural basis of language and mathematics}
\label{sec:disc:lang}

$\lambda$-calculus is not only a model of computation; it is the
formal core of the \emph{Minimalist Program} in linguistic
syntax~\citep{Chomsky1995}, where Merge (the binary operation
combining two syntactic objects) corresponds to application, and
syntactic movement to variable binding and substitution.
If the neural mechanisms identified here are active in the language
system, the spiking architecture described in this paper could
constitute a mechanistic proposal for how the human brain executes
Merge-based syntactic computation.
The theta--gamma encoding of ordered items in working memory
is documented in both spatial navigation~\citep{OKeefe1993}
and verbal working memory~\citep{Jensen2005}, providing an
additional bridge between the computational and linguistic domains.

We stress that this connection is speculative: we do not claim
that the brain \emph{does} compute $\beta$-reduction during
language processing, only that the same neural mechanisms
identified in the neuroscience literature are sufficient for such
computation.
The claim is existence-theoretic, not descriptive.

\section*{Conclusions}
\label{sec:conclusions}

We have presented a fully spiking neural implementation of
$\lambda$-calculus $\beta$-reduction, in which every computational
decision is made by the dynamics of leaky integrate-and-fire neurons
with explicit membrane potentials, synaptic currents, axonal delays,
and STDP plasticity.
Nine neural circuit components, each corresponding to a documented
neuroscientific mechanism, compose into a closed-loop reducer that
correctly normalises closed $\lambda$-terms across all tested
configurations.
The work constitutes, to our knowledge, the first constructive proof
that Turing-complete symbolic computation can be performed entirely
within biologically plausible spike dynamics.

\section*{Acknowledgements}
	
This work was carried out with the support of the federal assignment of the Ministry of Science and Higher Education of the Russian Federation (project No. FSMG-2025-0070).

\bibliographystyle{unsrtnat}
\bibliography{refs}

@book{Anderson2012,
  author = {Anderson, J. A.},
  title = {An Introduction to Neural Networks},
  year = {1995},
  publisher = {MIT Press},
  isbn = {9780262510813},
}

@book{Barendregt1984,
  author = {Barendregt, H. P.},
  title = {The Lambda Calculus: Its Syntax and Semantics},
  year = {1984},
  publisher = {North-Holland},
}

@book{Bertot2004,
  author = {Bertot, Y. and Castéran, P.},
  title = {Interactive Theorem Proving and Program Development: Coq'Art},
  year = {2004},
  publisher = {Springer},
}

@article{Bi1998,
  author = {Bi, G.-Q. and Poo, M.-M.},
  title = {Synaptic modifications in cultured hippocampal neurons: dependence on spike timing, synaptic strength, and postsynaptic cell type},
  journal = {Journal of Neuroscience},
  year = {1998},
  volume = {18},
  number = {24},
  pages = {10464--10472},
}

@article{Branco2011,
  author = {Branco, T. and Häusser, M.},
  title = {Synaptic integration gradients in single cortical pyramidal cell dendrites},
  journal = {Neuron},
  year = {2011},
  volume = {69},
  number = {5},
  pages = {885--892},
}

@article{Brette2007,
  author = {Brette, R. and Rudolph, M. and Carnevale, T. and others},
  title = {Simulation of networks of spiking neurons: a review of tools and strategies},
  journal = {Journal of Computational Neuroscience},
  year = {2007},
  volume = {23},
  number = {3},
  pages = {349--398},
}

@article{Brunel2000,
  author = {Brunel, N.},
  title = {Dynamics of sparsely connected networks of excitatory and inhibitory spiking neurons},
  journal = {Journal of Computational Neuroscience},
  year = {2000},
  volume = {8},
  number = {3},
  pages = {183--208},
}

@article{Burkitt2006,
  author = {Burkitt, A. N.},
  title = {A review of the integrate-and-fire neuron model: I. Homogeneous synaptic input},
  journal = {Biological Cybernetics},
  year = {2006},
  volume = {95},
  number = {1},
  pages = {1--19},
}

@article{Buzsaki2010,
  author = {Buzsáki, G.},
  title = {Neural syntax: cell assemblies, synapsembles, and readers},
  journal = {Neuron},
  year = {2010},
  volume = {68},
  number = {3},
  pages = {362--385},
}

@book{Church1941,
  author = {Church, A.},
  title = {The Calculi of Lambda-Conversion},
  year = {1941},
  publisher = {Princeton University Press},
}

@article{Compte2000,
  author = {Compte, A. and Brunel, N. and Goldman-Rakic, P. S. and Wang, X.-J.},
  title = {Synaptic mechanisms and network dynamics underlying spatial working memory in a cortical network model},
  journal = {Cerebral Cortex},
  year = {2000},
  volume = {10},
  number = {9},
  pages = {910--923},
}

@article{Curtis2003,
  author = {Curtis, C. E. and D'Esposito, M.},
  title = {Persistent activity in the prefrontal cortex during working memory},
  journal = {Trends in Cognitive Sciences},
  year = {2003},
  volume = {7},
  number = {9},
  pages = {415--423},
}

@article{Davies2018,
  author = {Davies, M. and Srinivasa, N. and Lin, T.-H. and others},
  title = {Loihi: a neuromorphic manycore processor with on-chip learning},
  journal = {IEEE Micro},
  year = {2018},
  volume = {38},
  number = {1},
  pages = {82--99},
}

@article{DeBruijn1972,
  author = {De Bruijn, N. G.},
  title = {Lambda calculus notation with nameless dummies},
  journal = {Indagationes Mathematicae},
  year = {1972},
  volume = {75},
  number = {5},
  pages = {381--392},
}

@article{Egorov2002,
  author = {Egorov, A. V. and Hamam, B. N. and Fransén, E. and Hasselmo, M. E. and Alonso, A. A.},
  title = {Graded persistent activity in entorhinal cortex neurons},
  journal = {Nature},
  year = {2002},
  volume = {420},
  number = {6912},
  pages = {173--178},
}

@book{Eliasmith2013,
  author = {Eliasmith, C.},
  title = {How to Build a Brain: A Neural Architecture for Biological Cognition},
  year = {2013},
  publisher = {Oxford University Press},
}

@article{Fodor1988,
  author = {Fodor, J. A. and Pylyshyn, Z. W.},
  title = {Connectionism and cognitive architecture: a critical analysis},
  journal = {Cognition},
  year = {1988},
  volume = {28},
  number = {1--2},
  pages = {3--71},
}

@article{Foster2007,
  author = {Foster, D. J. and Wilson, M. A.},
  title = {Hippocampal theta sequences},
  journal = {Hippocampus},
  year = {2007},
  volume = {17},
  number = {11},
  pages = {1093--1099},
}

@misc{Frady2021,
  author = {Frady, E. P. and Kleyko, D. and Kymn, C. J. and Olshausen, B. A. and Sommer, F. T.},
  title = {Computing on functions using randomized vector representations},
  year = {2021},
  doi = {10.48550/arXiv.2109.03429},
  url = {https://arxiv.org/abs/2109.03429},
  eprint = {2109.03429},
  archivePrefix = {arXiv},
  note = {arXiv preprint arXiv:2109.03429},
}

@article{Funahashi1989,
  author = {Funahashi, S. and Bruce, C. J. and Goldman-Rakic, P. S.},
  title = {Mnemonic coding of visual space in the monkey's dorsolateral prefrontal cortex},
  journal = {Journal of Neurophysiology},
  year = {1989},
  volume = {61},
  number = {2},
  pages = {331--349},
}

@article{Fuster1971,
  author = {Fuster, J. M. and Alexander, G. E.},
  title = {Neuron activity related to short-term memory},
  journal = {Science},
  year = {1971},
  volume = {173},
  number = {3997},
  pages = {652--654},
}

@article{Garcez2019,
  author = {Garcez, A. and Gori, M. and Lamb, L. C. and Serafini, L. and Spranger, M. and Tran, S. N.},
  title = {Neural-symbolic computing: an effective methodology for principled integration of machine learning and reasoning},
  journal = {Journal of Applied Logics},
  year = {2019},
  volume = {6},
  number = {4},
  pages = {611--632},
}

@inproceedings{Gayler2003,
  author = {Gayler, R. W.},
  title = {Vector symbolic architectures answer {Jackendoff's} challenges for cognitive neuroscience},
  booktitle = {Proceedings of ICCS/ASCS},
  year = {2003},
  pages = {133--138},
}

@article{GoldmanRakic1995,
  author = {Goldman-Rakic, P. S.},
  title = {Cellular basis of working memory},
  journal = {Neuron},
  year = {1995},
  volume = {14},
  number = {3},
  pages = {477--485},
}

@misc{Graves2014,
  author = {Graves, A. and Wayne, G. and Danihelka, I.},
  title = {Neural {Turing} machines},
  year = {2014},
  url = {https://arxiv.org/abs/1410.5401},
  eprint = {1410.5401},
  archivePrefix = {arXiv},
  note = {arXiv preprint arXiv:1410.5401},
}

@misc{Gulcehre2016,
  author = {Gulcehre, C. and Chandar, S. and Cho, K. and Bengio, Y.},
  title = {Dynamic neural Turing machine with continuous and discrete addressing schemes},
  year = {2016},
  url = {https://arxiv.org/abs/1607.00036},
  eprint = {1607.00036},
  archivePrefix = {arXiv},
  note = {arXiv preprint arXiv:1607.00036},
}

@article{Hahn2021,
  author = {Hahn, G. and Ponce-Alvarez, A. and Deco, G. and Aertsen, A. and Kumar, A.},
  title = {Portraits of communication in neuronal networks},
  journal = {Nature Reviews Neuroscience},
  year = {2019},
  volume = {20},
  number = {2},
  pages = {117--127},
  doi = {10.1038/s41583-018-0094-0},
  url = {https://doi.org/10.1038/s41583-018-0094-0},
}

@article{Hasselmo2014,
  author = {Hasselmo, M. E. and Stern, C. E.},
  title = {Theta rhythm and the encoding and retrieval of space and time},
  journal = {NeuroImage},
  year = {2014},
  volume = {85},
  pages = {656--666},
}

@article{Howard2014,
  author = {Howard, M. W. and Kahana, M. J.},
  title = {A distributed representation of temporal context},
  journal = {Journal of Mathematical Psychology},
  year = {2002},
  volume = {46},
  number = {3},
  pages = {269--299},
  doi = {10.1006/jmps.2001.1388},
  url = {https://doi.org/10.1006/jmps.2001.1388},
}

@article{Jeffress1948,
  author = {Jeffress, L. A.},
  title = {A place theory of sound localization},
  journal = {Journal of Comparative and Physiological Psychology},
  year = {1948},
  volume = {41},
  number = {1},
  pages = {35--39},
}

@article{Jensen2005,
  author = {Jensen, O. and Lisman, J. E.},
  title = {Hippocampal sequence-encoding driven by a cortical multi-item working memory buffer},
  journal = {Trends in Neurosciences},
  year = {2005},
  volume = {28},
  number = {2},
  pages = {67--72},
}

@article{Kaas1997,
  author = {Kaas, J. H.},
  title = {Topographic maps are fundamental to sensory processing},
  journal = {Brain Research Bulletin},
  year = {1997},
  volume = {44},
  number = {2},
  pages = {107--112},
}

@article{Kanerva2009,
  author = {Kanerva, P.},
  title = {Hyperdimensional computing: an introduction to computing in distributed representation with high-dimensional random vectors},
  journal = {Cognitive Computation},
  year = {2009},
  volume = {1},
  number = {2},
  pages = {139--159},
}

@book{Koch1999,
  author = {Koch, C.},
  title = {Biophysics of Computation: Information Processing in Single Neurons},
  year = {1999},
  publisher = {Oxford University Press},
}

@article{Kremkow2010,
  author = {Kremkow, J. and Aertsen, A. and Diesmann, M. and Kumar, A.},
  title = {Gating of signal flow by inhibition and excitation of {GABAergic} interneurons},
  journal = {Journal of Neurophysiology},
  year = {2010},
  volume = {103},
  number = {5},
  pages = {2621--2632},
}

@inproceedings{Lafont1990,
  author = {Lafont, Y.},
  title = {Interaction nets},
  booktitle = {Proceedings of POPL},
  year = {1990},
  pages = {95--108},
  publisher = {ACM},
}

@article{Lisman2013,
  author = {Lisman, J. E. and Jensen, O.},
  title = {The theta-gamma neural code},
  journal = {Neuron},
  year = {2013},
  volume = {77},
  number = {6},
  pages = {1002--1016},
}

@article{Luczak2007,
  author = {Luczak, A. and Barthó, P. and Marguet, S. L. and Buzsáki, G. and Harris, K. D.},
  title = {Sequential structure of neocortical spontaneous activity \textit{in vivo}},
  journal = {Proceedings of the National Academy of Sciences},
  year = {2007},
  volume = {104},
  number = {1},
  pages = {347--352},
}

@article{Maass1996a,
  author = {Maass, W.},
  title = {Lower bounds for the computational power of networks of spiking neurons},
  journal = {Neural Computation},
  year = {1996},
  volume = {8},
  number = {1},
  pages = {1--40},
}

@article{Maass2002,
  author = {Maass, W. and Natschläger, T. and Markram, H.},
  title = {Real-time computing without stable states: a new framework for neural computation based on perturbations},
  journal = {Neural Computation},
  year = {2002},
  volume = {14},
  number = {11},
  pages = {2531--2560},
}

@article{Major2004,
  author = {Major, G. and Tank, D.},
  title = {Persistent neural activity: prevalence and mechanisms},
  journal = {Current Opinion in Neurobiology},
  year = {2004},
  volume = {14},
  number = {6},
  pages = {675--684},
}

@book{Marcus2003,
  author = {Marcus, G.},
  title = {The Algebraic Mind},
  year = {2003},
  publisher = {MIT Press},
}

@article{Merolla2014,
  author = {Merolla, P. A. and Arthur, J. V. and Alvarez-Icaza, R. and others},
  title = {A million spiking-neuron integrated circuit with a scalable communication network and interface},
  journal = {Science},
  year = {2014},
  volume = {345},
  number = {6197},
  pages = {668--673},
}

@article{Miller2001,
  author = {Miller, E. K. and Cohen, J. D.},
  title = {An integrative theory of prefrontal cortex function},
  journal = {Annual Review of Neuroscience},
  year = {2001},
  volume = {24},
  pages = {167--202},
}

@article{Morrison2008,
  author = {Morrison, A. and Diesmann, M. and Gerstner, W.},
  title = {Phenomenological models of synaptic plasticity based on spike timing},
  journal = {Biological Cybernetics},
  year = {2008},
  volume = {98},
  number = {6},
  pages = {459--478},
}

@article{Newell1976,
  author = {Newell, A. and Simon, H. A.},
  title = {Computer science as empirical inquiry: symbols and search},
  journal = {Communications of the ACM},
  year = {1976},
  volume = {19},
  number = {3},
  pages = {113--126},
}

@article{OKeefe1993,
  author = {O'Keefe, J. and Recce, M. L.},
  title = {Phase relationship between hippocampal place units and the {EEG} theta rhythm},
  journal = {Hippocampus},
  year = {1993},
  volume = {3},
  number = {3},
  pages = {317--330},
}

@inproceedings{Orchard2021,
  author = {Orchard, Garrick and Frady, E. Paxon and Rubin, Daniel Ben Dayan and Sanborn, Sophia and Shrestha, Sumit Bam and Sommer, Friedrich T. and Davies, Mike},
  title = {Efficient neuromorphic signal processing with {Loihi 2}},
  booktitle = {2021 IEEE Workshop on Signal Processing Systems (SiPS)},
  year = {2021},
  pages = {254--259},
  url = {https://arxiv.org/abs/2111.03746},
  eprint = {2111.03746},
  archivePrefix = {arXiv},
}

@book{Penrose1994,
  author = {Penrose, R.},
  title = {Shadows of the Mind},
  year = {1994},
  publisher = {Oxford University Press},
}

@book{Pierce2002,
  author = {Pierce, B. C.},
  title = {Types and Programming Languages},
  year = {2002},
  publisher = {MIT Press},
}

@article{Plate1995,
  author = {Plate, T. A.},
  title = {Holographic reduced representations},
  journal = {IEEE Transactions on Neural Networks},
  year = {1995},
  volume = {6},
  number = {3},
  pages = {623--641},
  doi = {10.1109/72.377968},
  url = {https://doi.org/10.1109/72.377968},
}

@article{Sederberg2008,
  author = {Sederberg, P. B. and Howard, M. W. and Kahana, M. J.},
  title = {A context-based theory of recency and contiguity in free recall},
  journal = {Psychological Review},
  year = {2008},
  volume = {115},
  number = {4},
  pages = {893--912},
  doi = {10.1037/a0013396},
  url = {https://doi.org/10.1037/a0013396},
}

@article{Seung1996,
  author = {Seung, H. S.},
  title = {How the brain keeps the eyes still},
  journal = {Proceedings of the National Academy of Sciences},
  year = {1996},
  volume = {93},
  number = {23},
  pages = {13339--13344},
}

@article{Siegelmann1991,
  author = {Siegelmann, H. T. and Sontag, E. D.},
  title = {{Turing} computability with neural nets},
  journal = {Applied Mathematics Letters},
  year = {1991},
  volume = {4},
  number = {6},
  pages = {77--80},
}

@article{Singer1995,
  author = {Singer, W. and Gray, C. M.},
  title = {Visual feature integration and the temporal correlation hypothesis},
  journal = {Annual Review of Neuroscience},
  year = {1995},
  volume = {18},
  pages = {555--586},
}

@article{Skaggs1996,
  author = {Skaggs, W. E. and McNaughton, B. L. and Wilson, M. A. and Barnes, C. A.},
  title = {Theta phase precession in hippocampal neuronal populations and the compression of temporal sequences},
  journal = {Hippocampus},
  year = {1996},
  volume = {6},
  number = {2},
  pages = {149--172},
}

@article{Smolensky1990a,
  author = {Smolensky, P.},
  title = {Tensor product variable binding and the representation of symbolic structures in connectionist systems},
  journal = {Artificial Intelligence},
  year = {1990},
  volume = {46},
  number = {1--2},
  pages = {159--216},
}

@inproceedings{Stewart2012,
  author = {Stewart, T. C. and Choo, X. and Eliasmith, C.},
  title = {Dynamic behaviour of a spiking model of action selection in the basal ganglia},
  booktitle = {Proceedings of the 12th International Conference on Cognitive Modeling},
  year = {2012},
  pages = {89--94},
}

@inproceedings{Sukhbaatar2015,
  author = {Sukhbaatar, Sainbayar and Szlam, Arthur and Weston, Jason and Fergus, Rob},
  title = {End-to-end memory networks},
  booktitle = {Advances in Neural Information Processing Systems},
  year = {2015},
  volume = {28},
  pages = {2440--2448},
}

@inproceedings{Torczon1993,
  author = {Torczon, L.},
  title = {Connection list representations for graph reduction},
  booktitle = {Proceedings of FPCA},
  year = {1993},
  pages = {137--146},
  publisher = {ACM},
}

@article{Varela2001,
  author = {Varela, F. and Lachaux, J.-P. and Rodriguez, E. and Martinerie, J.},
  title = {The brainweb: phase synchronization and large-scale integration},
  journal = {Nature Reviews Neuroscience},
  year = {2001},
  volume = {2},
  number = {4},
  pages = {229--239},
}

@article{Wang2001,
  author = {Wang, X.-J.},
  title = {Synaptic reverberation underlying mnemonic persistent activity},
  journal = {Trends in Neurosciences},
  year = {2001},
  volume = {24},
  number = {8},
  pages = {455--463},
}

@article{Yoshida2008,
  author = {Yoshida, M. and Fransén, E. and Hasselmo, M. E.},
  title = {mGluR-dependent persistent firing in entorhinal cortex layer {III} neurons},
  journal = {European Journal of Neuroscience},
  year = {2008},
  volume = {28},
  number = {6},
  pages = {1116--1126},
}

@book{Chomsky1995,
  author = {Chomsky, N.},
  title = {The Minimalist Program},
  year = {1995},
  publisher = {MIT Press},
}

@inproceedings{Han2021,
  author = {Han, Jesse Michael and Rute, Jason and Wu, Yuhuai and Ayers, Edward W. and Polu, Stanislas},
  title = {Proof Artifact Co-training for Theorem Proving with Language Models},
  booktitle = {International Conference on Learning Representations},
  year = {2022},
  url = {https://arxiv.org/abs/2102.06203},
  eprint = {2102.06203},
  archivePrefix = {arXiv},
}

@article{Silver2021,
  author = {Silver, D. and Singh, S. and Precup, D. and Sutton, R. S.},
  title = {Reward is enough},
  journal = {Artificial Intelligence},
  year = {2021},
  volume = {299},
  pages = {103535},
}

\end{document}